Research Article

# Droplet Fusion as a Relaxation Process: Comparison with Shape Recovery of Newtonian and Viscoelastic Droplets

Mohammad Moein Naderi[1], Zhangli Peng[1,2,*], and Huan-Xiang Zhou[2,3,4,*]

[1]Department of Biomedical Engineering, University of Illinois Chicago, Chicago, IL 60607, USA
[2]Center for Bioinformatics and Quantitative Biology (CBQB), University of Illinois Chicago, Chicago, IL 60612, USA
[3]Department of Chemistry, University of Illinois Chicago, Chicago, IL 60607, USA
[4]Department of Physics, University of Illinois Chicago, Chicago, IL 60607, USA
[*]Correspondence: zhpeng@uic.edu, hzhou43@uic.edu

ABSTRACT Biomolecular condensates formed by phase separation inside cells often exhibit viscoelastic behavior, yet their shape recovery and fusion dynamics are frequently interpreted using purely viscous models. Here, we develop a unified theoretical and computational framework to quantify how viscoelasticity governs these two fundamental processes. We combine analytical theory for small-deformation shape recovery with axisymmetric finite-element simulations based on the Oldroyd-B constitutive model to systematically investigate both shape recovery and droplet fusion under comparable physical conditions. Our results show that, although both processes are driven by capillary forces, they are fundamentally distinct in their underlying physics. Shape recovery is governed by global viscocapillary relaxation of a single connected interface and follows single- or multi-exponential decay depending on the relative magnitude of the viscocapillary timescale and the stress relaxation time. In contrast, droplet fusion is intrinsically a multistage process involving localized curvature-driven neck formation, rapid bridge expansion, and a transition to global relaxation. We demonstrate that viscoelasticity introduces an additional intrinsic timescale that governs the competition between capillary driving and stress relaxation, characterized by the Deborah number. This leads to enhanced intermediate-stage fusion dynamics and modified relaxation behavior compared to Newtonian droplets. Furthermore, we show that the presence of an exterior fluid with finite viscosity introduces additional hydrodynamic dissipation through two-phase coupling, significantly slowing the fusion process. Finally, we compared the computationally predicted droplet fusion in the Newtonian and viscoelastic cases with a stretched-exponential empirical formula. Deviations observed in viscoelastic regimes highlight the limitations of purely viscous descriptions and the need for models incorporating stress relaxation and memory effects.

SIGNIFICANCE We develop a unified theoretical and computational framework to compare droplet shape recovery and fusion under comparable conditions. Although both processes are driven by capillary forces, we show that their underlying physics are fundamentally different: shape recovery is governed by global viscocapillary relaxation, whereas droplet fusion proceeds through multiple stages involving localized neck formation and non-exponential dynamics. Viscoelasticity introduces an additional intrinsic timescale that alters relaxation behavior and enhances intermediate-stage fusion dynamics. These findings provide a mechanistic framework for interpreting deformation and fusion experiments and for extracting rheological properties of biomolecular condensates.

## INTRODUCTION

Biomolecular condensates are formed through a phase separation process, in which an initially homogeneous liquid is divided into a dense, macromolecule-rich phase and a surrounding dilute phase (1–6). The material properties of many condensates evolve over time, with a progressive shift toward more solid-like behavior, a phenomenon commonly referred to as “aging" (7–13). Such condensates play an important role in various biological and pathological processes, including cancer (14, 15) and neurodegenerative diseases (16), and cellular stress responses (17). As an example, phase separation contributes to cancer by altering key regulatory programs that drive tumor development (3).

These phase-separated condensates, often referred to as condensate droplets or simply droplets, exhibit behaviors that reflect their underlying material properties. Two fundamental processes that govern the shape dynamics of these droplets are shape recovery (relaxation) and fusion (Fig. 1). In shape recovery, an initially deformed droplet relaxes into a spherical shape, while in fusion, two droplets are merged together and form a single droplet. Both processes lead to the minimization of surface energy driven by capillary forces.

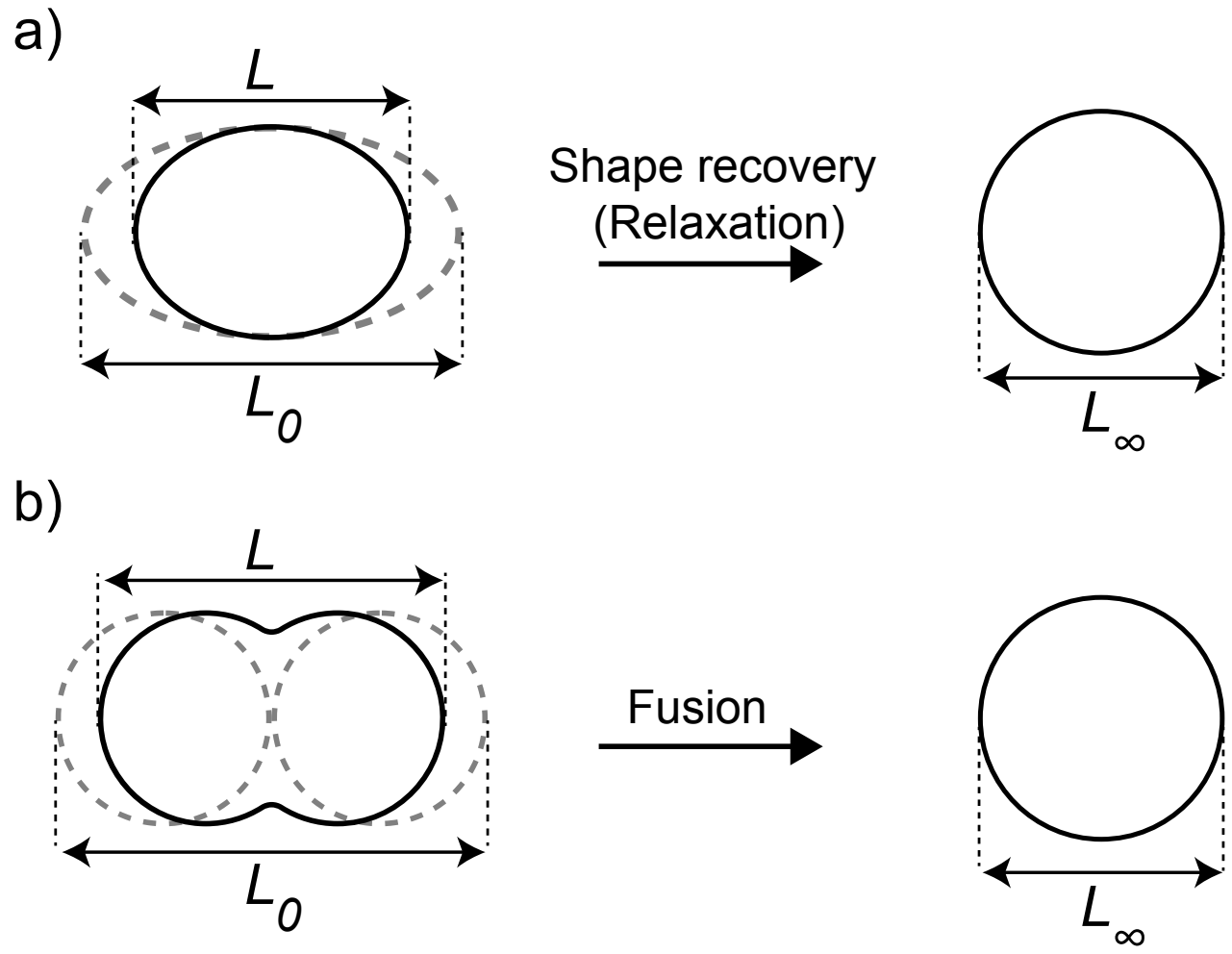

Figure 1: Schematic comparison of droplet shape recovery and droplet fusion. a) shape recovery (relaxation) of an initially deformed droplet. b) The fusion of two droplets coalescing into a single droplet. $L_0$, $L$, and $L_\infty$ are the initial, instantaneous, and final length of the droplet.

Over the years, biomolecular droplets have commonly been modeled as purely viscous liquids, characterized by instantaneous shear stress relaxation and Newtonian rheology. Within this framework, early experimental studies by Brangwynne et al. (13) and Elbaum-Garfinkle (18) interpreted droplet shape recovery and fusion dynamics as being governed solely by a balance between viscosity $\eta$ and surface tension $\gamma$, with a characteristic timescale set by the viscocapillary ratio, $\tau_{vc} = \eta R/\gamma$, where $R$ is the droplet radius. More recently, optical-tweezers experiments by Ghosh et al. (19) and other studies (20–23) demonstrated that biomolecular condensates cannot, in general, be described as purely viscous fluids during droplet fusion. Instead, these condensates exhibit pronounced viscoelastic behavior, implying finite shear stress relaxation times that are comparable to or longer than the viscocapillary timescale. As a result, the interplay between stress relaxation and capillary forces can substantially alter the fusion dynamics, leading to behavior that cannot be captured by Newtonian models alone.

Prosperetti (24, 25) presented a full analytical solution to the recovery dynamics of viscous droplets from a small-amplitude deformation. Phenomenological models have been developed to describe the shape recovery (deformation retraction) of viscoelastic droplets (26–28). This problem has also been investigated through direct numerical solutions of the governing fluid dynamics equations. In addition, numerous experimental studies on the deformation retraction of polymer blends (29, 30) have been conducted, consistently highlighting the significant role of viscoelastic effects in governing relaxation dynamics. Khismatullin and Nadim (31) derived the normal mode solution of shape oscillations of a viscoelastic drop explicitly for the Jeffreys model of viscoelasticity.

In a recent theoretical study, Zhou (22) derived an exact analytical solution for the shape recovery of deformed viscoelastic droplets in the time domain under the assumption of small-amplitude deformations in the Stokes regime. This work provided the first analytical description of how a freely relaxing viscoelastic droplet returns to its spherical shape after deformation, and studied the effect of the external fluid domains. The analysis revealed that viscoelastic droplets generally exhibit multi-exponential recovery dynamics, controlled jointly by the viscocapillary timescale ($\tau_{vc}$) and the intrinsic shear relaxation time ($t_c$). A Deborah number can be defined as $De = t_c/\tau_{vc}$. Depending on the relative magnitude of these timescales, viscoelastic droplets may display apparent shear thickening at fast relaxation rates or shear thinning at slow relaxation rates—phenomena absent in purely viscous models.

Despite this progress on shape recovery, no analytical solution currently exists for droplet fusion, either in the viscous or viscoelastic regime, owing to the inherently complex and strongly nonlinear two-droplet geometry during coalescence. In the absence of such a theory, Ghosh et al. and Zhou (19, 32) proposed an empirical stretched-exponential form $1 - e^{-(t/\tau_f)^\beta}$, where $\beta$ is the stretching constant and $\tau_f$ is one characteristic relaxation time, to fit experimental fusion trajectories, but the theoretical basis and optimality of this functional form remain unclear. Zhou (22) further compared fusion and shape-recovery dynamics within the viscous framework and argued that fusion proceeds more slowly than shape recovery, with fusion following a stretched-exponential time dependence, whereas shape recovery accelerates with increasing deformation mode number (e.g., higher-order Legendre modes). However, viscous theory alone was shown to be insufficient to account for the orders-of-magnitude separation between fusion and recovery times observed experimentally in some systems, suggesting an essential role for viscoelasticity.

In this work, we address these open questions by performing finite-element simulations of both droplet fusion and droplet shape recovery under comparable conditions using a viscoelastic constitutive model. By systematically varying viscosity, polymeric stress, and relaxation time, we directly compare the two classes of shape change within a unified computational framework. Our goal is to elucidate both the shared physical mechanisms and the fundamental differences in the dynamics of fusion and recovery in viscoelastic droplets, thereby bridging the gap between analytical theory and experimental observations.

# METHODS

## Analytical model for shape recovery of a viscoelastic droplet

We developed an analytical theory of the shape recovery of a viscoelastic droplet (32). A brief summary of the theory and its extension to the droplet fusion process are given next. We consider the relaxation of a weakly deformed, axisymmetric droplet of radius $R$ embedded in a fluid of negligible inertia. The droplet interface is represented as a perturbation about the spherical shape,

$$r(\theta, t) = R + f_\ell(t) P_\ell(\cos\theta), \tag{1}$$

where $P_\ell$ is the Legendre polynomial of degree $\ell \geq 2$ and $f_\ell(t)$ denotes the time-dependent deformation amplitude.

### Purely viscous (Newtonian) droplet

For a Newtonian droplet with viscosity $\eta$, the linearized hydrodynamic equations yield an exponential recovery of each deformation mode,

$$f_\ell(t) = f_\ell(0)\, e^{-\lambda_\ell^{(D)} t}, \tag{2}$$

with relaxation rate

$$\lambda_\ell^{(D)} = \frac{\ell(\ell+2)(2\ell+1)}{2(2\ell^2+4\ell+3)} \frac{\gamma}{\eta R}, \tag{3}$$

where $\gamma$ is the interfacial tension. This relaxation is governed solely by the viscocapillary timescale $\tau_{vc} = \eta R/\gamma$.

### Viscoelastic droplet

To incorporate viscoelasticity, the droplet material is described by a linear viscoelastic constitutive relation with shear relaxation modulus $G(t)$. In Laplace space, the Newtonian viscosity $\eta$ is replaced by the Laplace-transformed modulus $\hat{G}(s)$, yielding

$$\hat{f}_\ell(s) = \frac{f_\ell(0)}{s + \lambda_\ell^{(D)}/\hat{g}(s)}, \tag{4}$$

where

$$\hat{g}(s) = \frac{\hat{G}(s)}{\eta_z}, \tag{5}$$

and $\eta_z = \hat{G}(0)$ is the zero-shear viscosity.

### Jeffreys viscoelastic model

For a Jeffreys fluid, the shear relaxation modulus is

$$\hat{G}(s) = \eta_0 + \frac{\eta_p}{1 + t_c s}, \tag{6}$$

where $\eta_0$ is the Newtonian solvent viscosity, $\eta_p$ is the polymeric viscosity, $\eta_z = \eta_0 + \eta_p$ is the total zero shear viscosity, and $t_c$ is the stress relaxation time. The shape recovery becomes bi-exponential,

$$f_\ell(t) = f_\ell(0)\left(A_{\ell,+}e^{-\lambda_{\ell,+}t} + A_{\ell,-}e^{-\lambda_{\ell,-}t}\right), \tag{7}$$

with relaxation rates

$$\lambda_{\ell,\pm} = \frac{1}{2}\frac{\eta_z}{\eta_0}\frac{1}{t_c}\left[t_c\lambda_\ell^{(D)} + 1 \pm \sqrt{\left(t_c\lambda_\ell^{(D)} + 1\right)^2 - 4\frac{\eta_0}{\eta_z}t_c\lambda_\ell^{(D)}}\right], \tag{8}$$

and corresponding amplitudes

$$A_{\ell,\pm} = \frac{1}{2}\left[1 \pm \frac{t_c\lambda_\ell^{(D)} - 1}{\sqrt{\left(t_c\lambda_\ell^{(D)} + 1\right)^2 - 4\frac{\eta_0}{\eta_z}t_c\lambda_\ell^{(D)}}}\right]. \tag{9}$$

### Physical interpretation

The viscoelastic recovery dynamics is governed by the competition between the viscocapillary timescale $\tau_{vc}$ and the shear relaxation time $t_c$. When $t_c\lambda_\ell^{(D)} \ll 1$, the droplet behaves effectively as a Newtonian fluid with single-exponential relaxation. When $t_c\lambda_\ell^{(D)} \gtrsim 1$, viscoelastic stress relaxation introduces additional timescales, leading to multi-exponential recovery, apparent shear thickening at fast relaxation, and shear thinning at slow relaxation.

## Computational Modeling

We applied the finite element software package COMSOL Multiphysics (33) (COMSOL, Inc., Stockholm, Sweden) to simulate the shape recovery of an elongated droplet and the fusion of two droplets. We modeled the domains as axisymmetric configurations in all cases. We applied the moving mesh interface based on the arbitrary Lagrangian Eulerian (ALE) formulation (34) to model the motion of the interface and the deformation of the meshes. The Yeoh hyperelastic model (35) was employed to describe the deforming mesh behavior. An axisymmetric mesh consisting of first-order triangular elements was used to discretize the domains, and convergence was tested using different resolutions. The consistent streamline stabilization method was applied. The inertial forces were neglected since the Reynolds number is very small in all cases. A Backward Differentiation Formula (BDF) (36) scheme was employed for time integration, and the Newton method was applied to handle the nonlinear equations. The MUMPS direct solver (37) was applied to solve the linear equations. A parametric sweep was conducted to study the effects of relaxation time and viscosity of the droplets on the processes.

### Viscoelastic Droplet Model

We utilized the Viscoelastic Flow (*vef*) interface in COMSOL Multiphysics to simulate the droplet shape recovery and fusion processes. The governing equations consist of the incompressible Navier-Stokes equation (38):

$$\rho\left(\frac{\partial \mathbf{u}}{\partial t} + \mathbf{u}\cdot\nabla\mathbf{u}\right) = -\nabla p + \eta_s\nabla^2\mathbf{u} + \nabla\cdot\boldsymbol{\tau}_p, \quad \nabla\cdot\mathbf{u} = 0 \tag{10}$$

where $\mathbf{u}$ denotes the velocity field, $p$ the pressure, $\eta_s$ the solvent viscosity, and $\boldsymbol{\tau}_p$ the polymeric stress tensor. $\rho$ is the density of the fluid, but due to the small length scale, the inertial terms on the left side of Eq. 10 are ignored.

The Oldroyd-B model (39) was used as the constitutive equation, which is known to be difficult for numerical implementation and to be prone to numerical instability (40–42). Note that Jeffreys viscoelastic model is the linear limit of Oldroyd-B model. The polymeric viscosity ($\eta_p$) and relaxation time ($t_c$) were prescribed as model parameters to represent the viscoelastic behavior of the droplets.

$$\boldsymbol{\tau}_p + t_c\overset{\triangledown}{\boldsymbol{\tau}}_p = 2\eta_p\mathbf{D}, \tag{11}$$

where the upper-convected derivative is defined as:

$$\overset{\triangledown}{\boldsymbol{\tau}}_p = \frac{\partial\boldsymbol{\tau}_p}{\partial t} + \mathbf{u}\cdot\nabla\boldsymbol{\tau}_p - (\nabla\mathbf{u})^T\cdot\boldsymbol{\tau}_p - \boldsymbol{\tau}_p\cdot(\nabla\mathbf{u}), \tag{12}$$

$$\mathbf{D} = \frac{1}{2}\left[\nabla\mathbf{u} + (\nabla\mathbf{u})^T\right], \tag{13}$$

is the rate-of-deformation tensor. Please note that throughout this paper, the total viscosity of the droplet is fixed at $\eta_p + \eta_s = 0.5$ Pa·s for all cases.

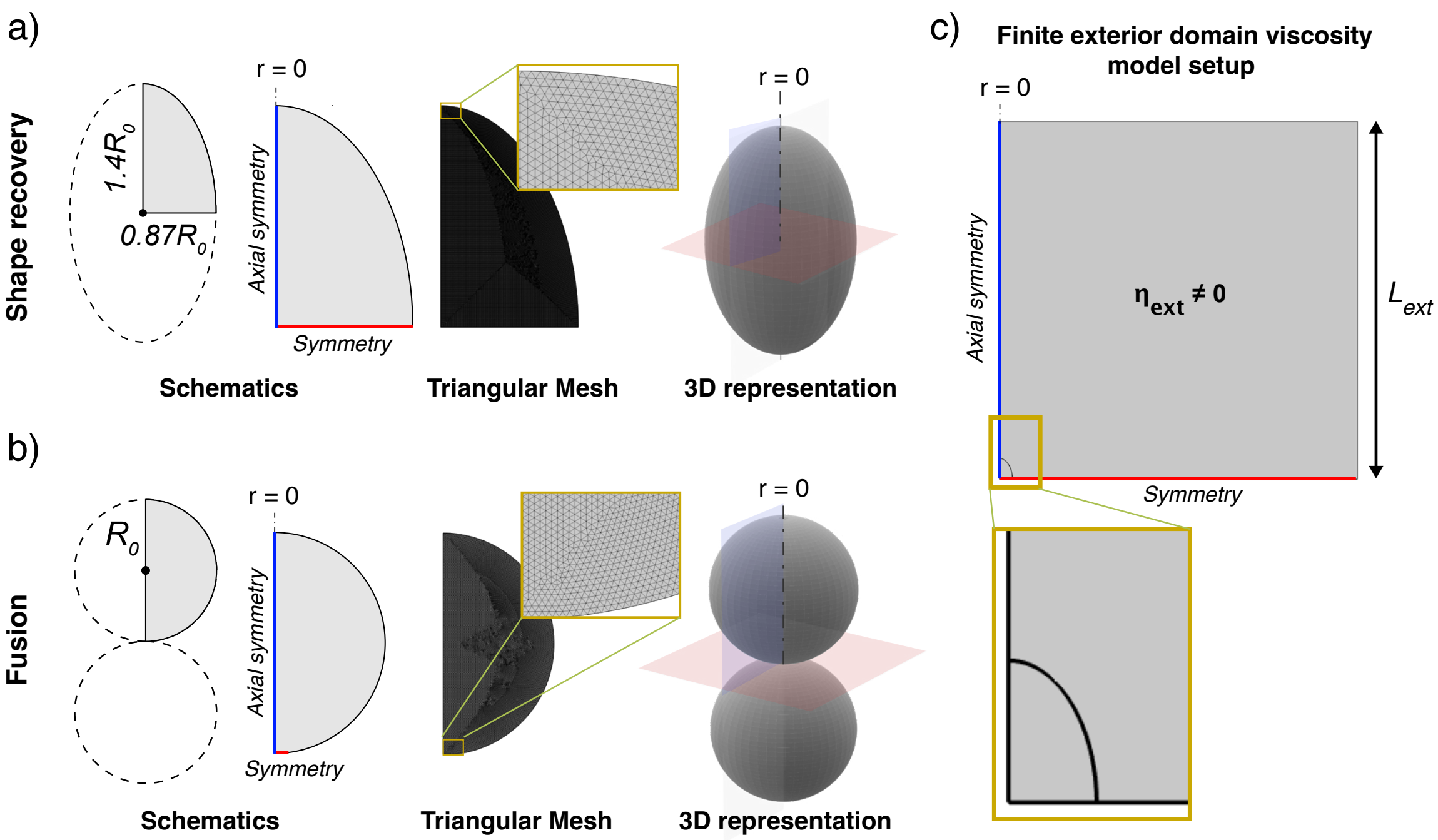

Figure 2: Numerical model schematics and mesh configuration for a) shape recovery, and b) fusion in an inviscid exterior domain, and c) shape recovery within an exterior domain with finite viscosity $\eta_{ext}$. Appropriate symmetry and axial symmetry boundary conditions were used to reduce computational cost.

## Boundary Conditions

The moving mesh interface was used, with the surface tension force included in the momentum equations. In cases without an exterior domain, the moving mesh interface represents the free surface using the *Free Surface* interface:

$$\boldsymbol{n}\cdot\boldsymbol{T} = -p_{\text{ext}}\boldsymbol{n} + \gamma(\nabla_t\cdot\boldsymbol{n})\boldsymbol{n} - \nabla_t\gamma, \tag{14}$$

where $\boldsymbol{n}$ is the unit normal vector, $p_{\text{ext}}$ is the external pressure, $\gamma$ is the surface tension coefficient, $\nabla_t$ denotes the tangential (surface) gradient operator, with the total stress tensor defined as:

$$\boldsymbol{T} = -p\mathbf{I} + \mathbf{K} + \boldsymbol{\tau}_p, \tag{15}$$

where $\mathbf{K} = 2\eta_s\mathbf{D}$ is the viscous stress tensor.

For cases with finite exterior viscosity, the moving mesh defines the interface between the internal and external domains using the *Fluid-Fluid* interface:

$$\boldsymbol{n}\cdot(\boldsymbol{T}_1 - \boldsymbol{T}_2) = \gamma\,(\nabla_t\cdot\boldsymbol{n})\,\boldsymbol{n} - \nabla_t\gamma, \tag{16}$$

where $\boldsymbol{T}_1$ and $\boldsymbol{T}_2$ are the total stress tensors of the interior and exterior fluids, respectively.

For the cases with finite exterior viscosity, a no-slip wall boundary condition was applied on the exterior walls, and the exterior domain is modeled large enough to minimize the effects of these stationary walls on the particle-fluid interaction. In all cases, *Axial Symmetry* boundary condition is used on the longer axis of the droplet(s), and *Symmetry* boundary condition is applied to the shorter axis. Consequently, as shown in Fig. 2, with the appropriate symmetry boundary conditions, the full 3-D elongated ellipsoidal droplet in the shape recovery cases, and the pair of 3-D droplets in fusion cases, were initially represented as a quarter of a 2-D oval and a semicircle, respectively. This significantly reduces the computational cost of the simulations.

## Mesh Independence and Model Validation

A mesh refinement convergence study was performed to ensure that the simulation results are independent of the element size. For this purpose, we plotted $\frac{L(t)-L(\infty)}{L(0)-L(\infty)}$ as a function of time for a shape recovery of a deformed droplet for different resolutions of meshes, where $L$ denotes the instantaneous length of the droplet, and $L_0$ and $L_\infty$ represent the initial and final length of the droplet, respectively. Note that for the shape-recovery case, the droplet's initial radii along the short and long axes are set to $0.87R_0$ and $1.4R_0$, respectively. This yields an initial aspect ratio of 1:1.6. This is the fixed initial aspect ratio of the deformed droplet for all cases in this paper, unless otherwise stated. Fig. 3 shows that as the number of elements were increased from 25,000 to 70,000, mesh independence was observed.

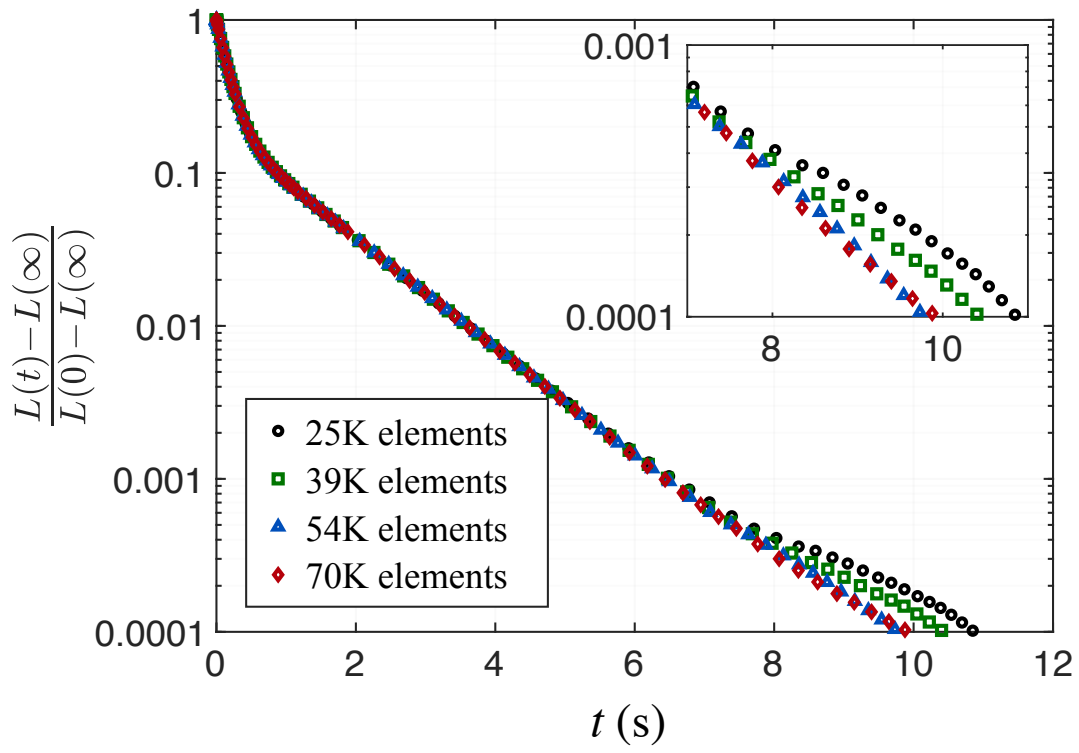

Figure 3: Mesh independence study showing the effect of the number of elements on the droplet shape recovery.

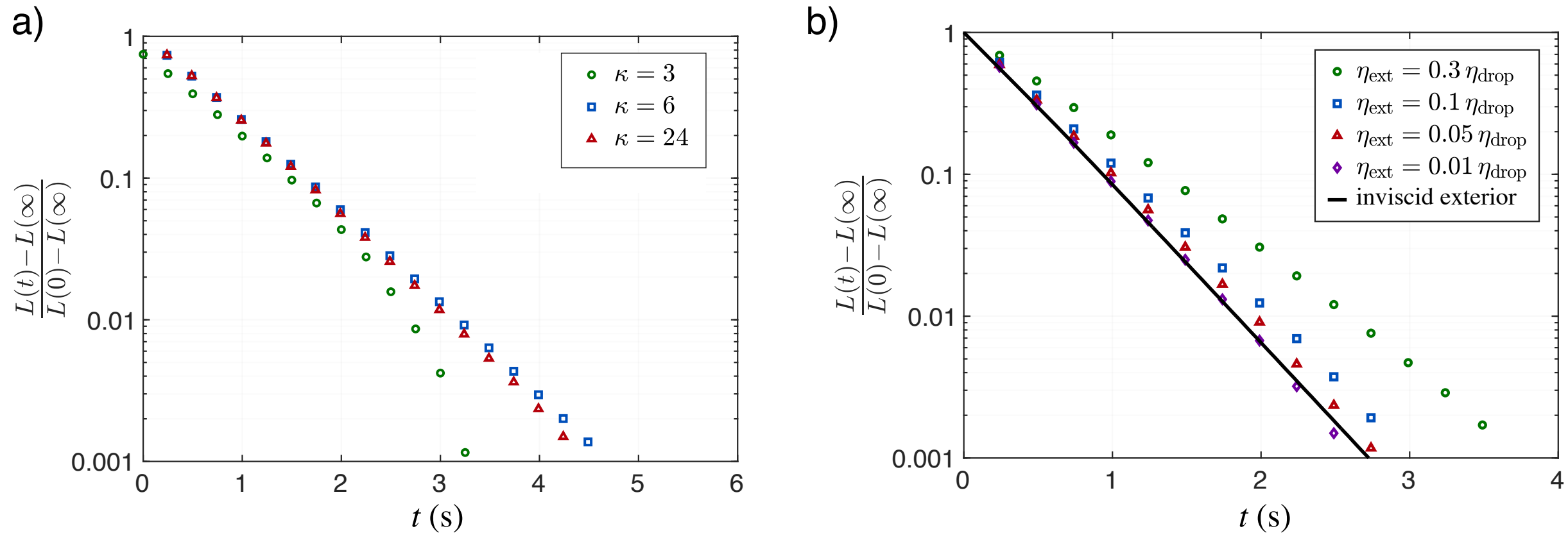

Figure 4: Effects of exterior domain. (a) Effect of domain confinement ratio ($\kappa$) on the accuracy of the exterior domain model. (b) Validation of the cases with finite viscosity exterior domain.

For the droplet fusion and exterior domain models, a different validation strategy was employed. Due to the higher complexity of these cases, refining the mesh simply by increasing the number of elements often led to convergence issues. Therefore, to ensure convergence with the highest possible accuracy, various combinations of mesh density and domain size (for exterior domain models) were examined. The most accurate configuration was identified by systematically monitoring the simulation results. The dependence of the results on the size of the exterior domain was investigated first. Three values for the exterior domain confinement ratio $\kappa = L_{ext}/R_0$ (the ratio of the exterior domain length to the initial droplet radius) were tested. As expected, the results converged by increasing the confinement ratio ($\kappa$), as illustrated in Fig. 4a. Next, the viscosity of the exterior domain was reduced incrementally and the behavior of the droplet was monitored. As shown in Fig. 4b, the simulation results converged to that of the inviscid exterior case as the exterior viscosity approached zero, providing additional confirmation of the validity of the model.

## RESULTS

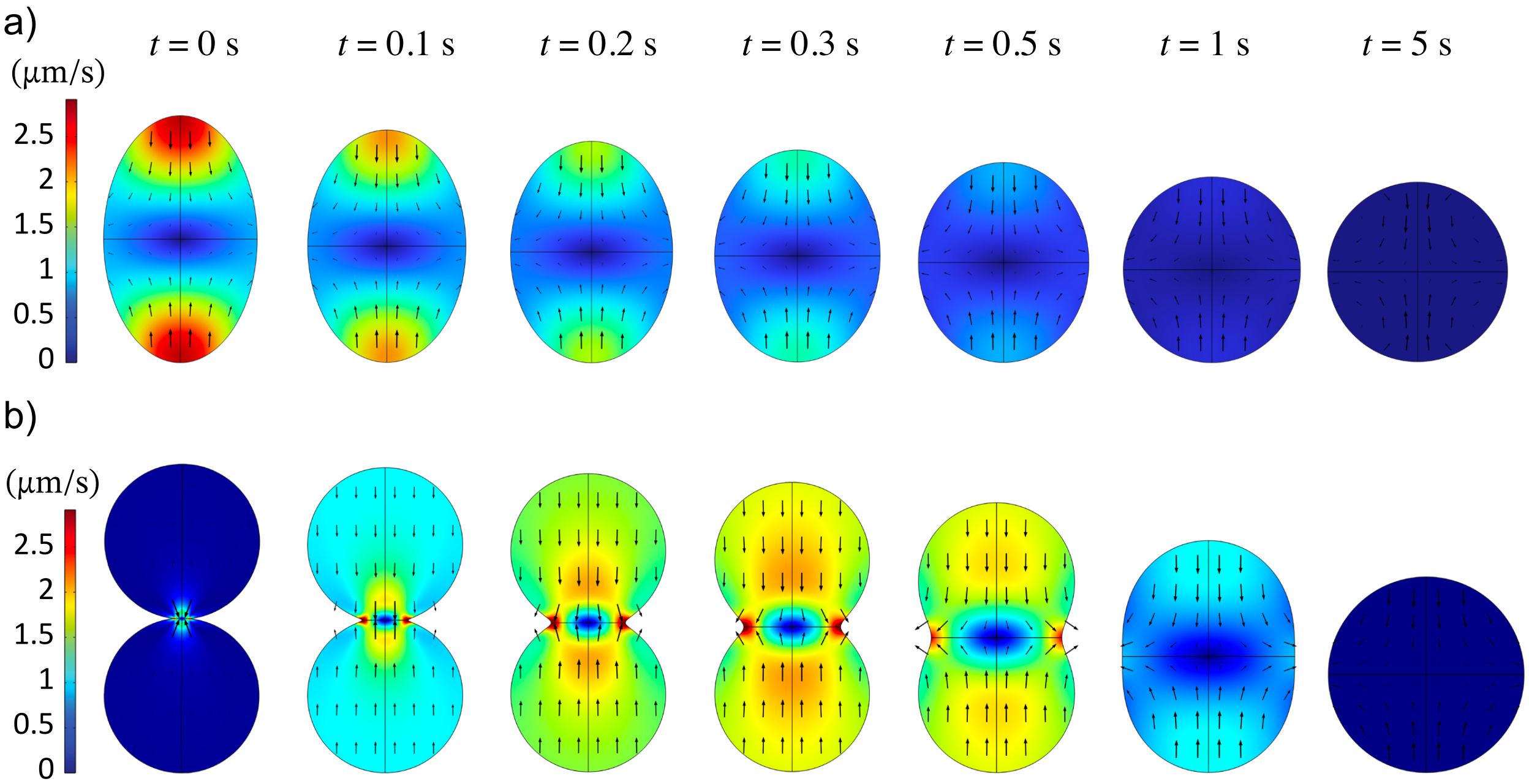

Figure 5: Comparison of shape recovery (a) and droplet fusion (b) dynamics for Newtonian droplets. Contours show the velocity on the mid-plane cross-section. The viscosity and the surface tension of the droplet(s) are fixed at $\eta = 0.5\,\text{Pa·s}$ and $\gamma = 6 \times 10^{-6} N/m$.

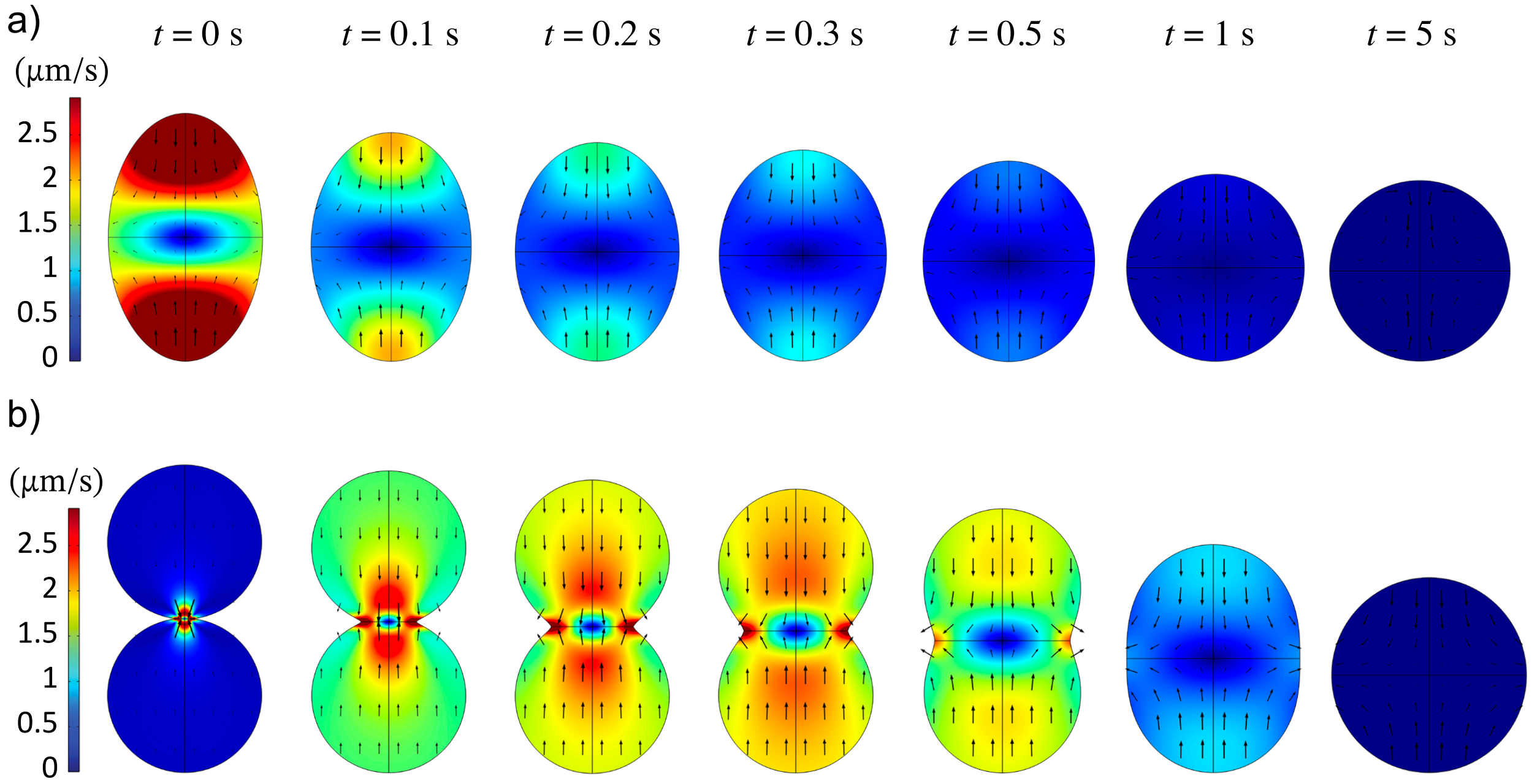

Figure 6: Comparison of shape recovery (a) and droplet fusion (b) dynamics for a viscoelastic droplet. Contours show the velocity on the mid-plane cross-section. The polymeric viscosity and relaxation time of the droplet are fixed at $\eta_p = 0.25\,\text{Pa·s}$ and $t_c = 0.1$ s. Total viscosity is $\eta = \eta_p + \eta_s = 0.5\,\text{Pa·s}$. The surface tension is fixed at $\gamma = 6 \times 10^{-6} N/m$.

First, we present a systematic visualization of the computationally predicted deformation histories and associated velocity fields for both droplet shape recovery and droplet fusion. Second, we validate our numerical framework by comparing the simulated shape recovery dynamics against available analytical solutions, covering both small-deformation regimes—where linear theory applies—and large-deformation cases—where nonlinear effects become significant. Third, we perform a detailed, side-by-side quantitative comparison of shape recovery and fusion dynamics. By examining key metrics such as characteristic length scales, relaxation rates, and velocity evolution, we identify both the common physical mechanisms and the fundamental differences between the two processes. Finally, we evaluate the ability of existing empirical models, developed primarily for Newtonian droplets, to describe the fusion process.

### Comparison of deformation history and velocity fields between shape recovery and fusion

We examine the computationally predicted droplet shapes and corresponding velocity fields for two canonical processes: the shape recovery of an initially elongated Newtonian droplet (Fig. 5a) and the fusion of two initially spherical Newtonian droplets (Fig. 5b). In both cases, the dynamics are governed by a balance between capillary forces arising from interfacial curvature and viscous stresses within the fluid, with inertia being negligible. The velocity fields therefore directly reflect how curvature gradients drive fluid motion to minimize surface energy.

For shape recovery, the flow is relatively simple and can be interpreted within the framework of capillary-driven relaxation of a single connected interface. The elongated droplet initially exhibits strong curvature variation, with the highest mean curvature located at the two ends. This curvature gradient generates a capillary pressure difference that drives fluid from the high-curvature tips toward the lower-curvature region. Consequently, the largest velocities are observed near the ends at early times. The flow field is smooth and globally coordinated, with no topological change in the interface. In the interior region, the velocity field resembles a uniaxial extensional flow, characterized by axial compression and radial expansion, consistent with the redistribution of volume required to approach a sphere. As the droplet relaxes, curvature differences diminish, leading to a monotonic decay of both velocity magnitude and deformation amplitude. This behavior is consistent with classical viscocapillary relaxation, where a single dominant timescale emerges from the ratio of viscosity to surface tension.

In contrast, droplet fusion involves a fundamentally different geometric and dynamical structure due to the initial presence of two disconnected interfaces and the subsequent topological transition. The process can be divided into several stages. In the initial stage (approximately 0–0.1 s), the droplets come into contact at a point, creating a singular geometry with formally divergent curvature. This results in a localized region of extremely high capillary pressure, which drives the rapid formation of a liquid bridge between the droplets. In the second stage (approximately 0.2–0.5 s), the bridge expands under the action of this strong capillary pressure. The driving force is dominated by the high curvature of the neck region, leading to a rapid increase in bridge radius. Despite this strong local flow, the exact center of the neck remains a stagnation point due to symmetry, as evidenced by the near-zero velocity region in the contour plots. The flow field in this regime is highly localized and exhibits strong spatial gradients, in contrast to the global flow observed in shape recovery. In the third stage (after approximately 1 s), as the neck radius increases and the interface becomes smoother, the curvature gradients decrease and the flow transitions from a localized neck-driven regime to a more global relaxation. Once the neck becomes convex, the merged droplet behaves as a single connected body, and the dynamics resemble those of shape recovery, governed by large-scale redistribution of fluid toward a spherical equilibrium. Thus, droplet fusion is inherently a multiscale process, transitioning from a singular, curvature-dominated regime to a global viscocapillary relaxation.

For viscoelastic droplets (Fig. 6), the overall geometric evolution remains qualitatively similar, but the underlying stress balance is modified by the presence of elastic stresses and finite relaxation times. In addition to viscous dissipation, the fluid now stores elastic energy, which can be released during deformation. In the case of shape recovery, this leads to a stronger initial driving force, as both capillary pressure and elastic stress contribute to the motion. As a result, higher velocity magnitudes are observed at early times ($t = 0$ s in Fig. 6a). The subsequent evolution reflects the competition between capillary relaxation and stress relaxation, which can introduce additional timescales beyond the classical viscocapillary time.

Similarly, in droplet fusion, viscoelasticity modifies the dynamics of bridge expansion. The presence of elastic stress enhances the effective driving force near the neck, leading to higher velocities and a faster initial growth of the bridge (Fig. 6b). At the same time, stress relaxation can redistribute stored elastic energy over time, altering the temporal evolution of the flow field compared to the Newtonian case. These effects are particularly pronounced in the early and intermediate stages, where curvature gradients and deformation rates are largest. Overall, the velocity field provides direct evidence that viscoelasticity not only changes the magnitude of the flow but also modifies the interplay between local curvature-driven dynamics and global relaxation, even when the macroscopic shape evolution appears similar.

### Comparison of shape recovery simulations with the analytical theory

Next, we quantitatively compare the analytical theory of shape recovery with the numerical simulation results. Figure 7 presents this comparison between finite-element simulations based on the Oldroyd-B constitutive model and the analytical solution derived from the Jeffreys viscoelastic model. Two representative initial geometries are considered: (a) a moderately deformed droplet with an initial aspect ratio of 1:1.6, corresponding to the small-deformation regime, and (b) a highly elongated droplet with an aspect ratio of 1:8, representing a strongly nonlinear deformation. The results are shown for both purely viscous (Newtonian) droplets and viscoelastic droplets with fixed polymeric viscosity $\eta_p$ = 0.25Pa·s, while varying the stress relaxation time over the range $t_c$ = 0.1–1 s.

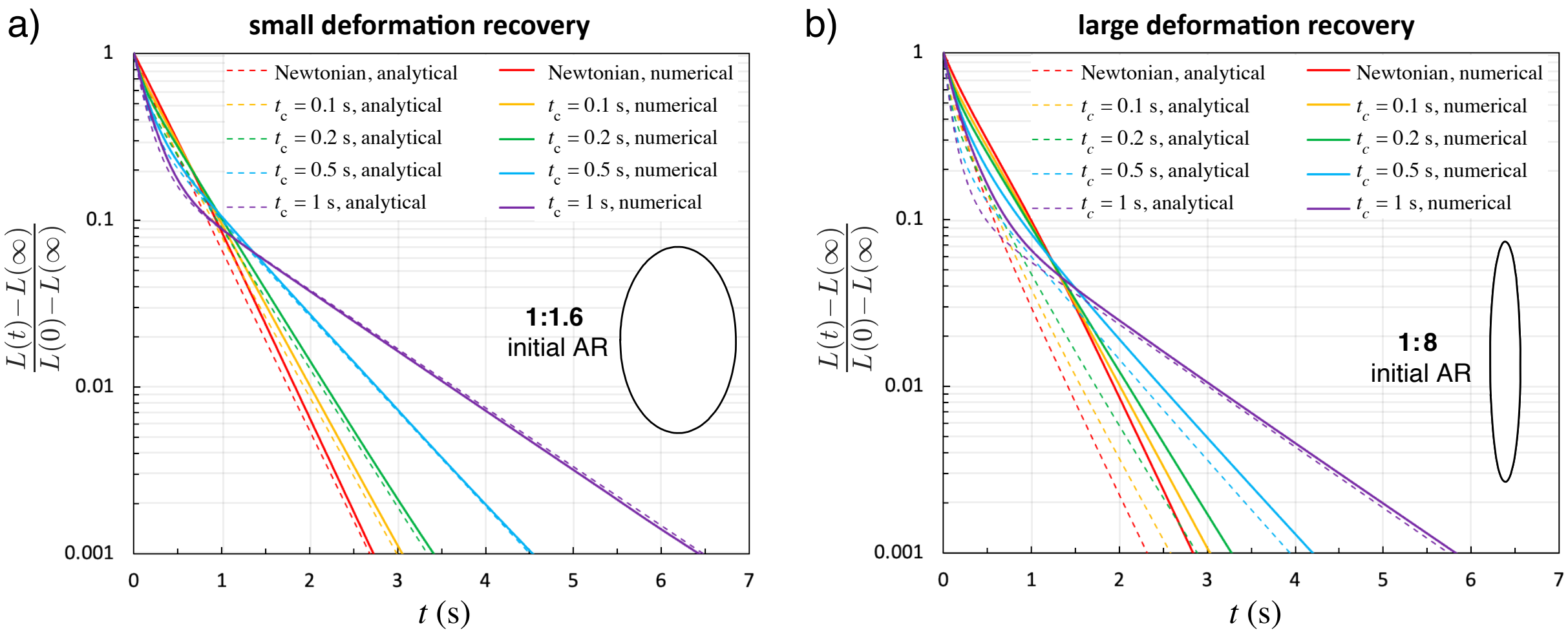

Figure 7: Comparison between the simulation results and the analytical solution for the shape recovery of a droplet with initial aspect ratios (AR) of (a) 1:1.6 (small deformation) and (b) 1:8 (large deformation). The polymeric viscosity of the viscoelastic droplet is fixed at $\eta_p$ = 0.25 Pa·s for both cases and total viscosity is $\eta$ = 0.5 Pa·s

From a theoretical standpoint, the analytical solution is constructed by linearizing the interfacial dynamics around a spherical shape and decomposing the interface perturbation into a sum of spherical harmonic modes. Each mode relaxes independently with a characteristic rate determined by the balance between capillary pressure (set by interfacial curvature) and viscous or viscoelastic resistance. In the Newtonian limit, this leads to a single exponential decay governed by the viscocapillary timescale $\tau_{vc} \sim \eta R/\gamma$. In contrast, for viscoelastic droplets described by the Jeffreys model, the coupling between capillary forces and stress relaxation introduces additional timescales, resulting in multi-exponential decay behavior controlled by both $\tau_{vc}$ and the intrinsic relaxation time $t_c$.

For the small-deformation case (Fig. 7a), the agreement between analytical and numerical results is excellent across all cases. This is expected because the assumptions underlying the analytical theory—namely, small interface perturbations, linearization of curvature, and decoupling of deformation modes—are well satisfied. In this regime, the droplet shape can be accurately described by a small number of low-order modes, and nonlinear geometric effects such as mode coupling and finite-amplitude curvature corrections are negligible. As a result, both the Newtonian and viscoelastic simulations closely follow the predicted exponential or multi-exponential relaxation dynamics.

In the large-deformation case (Fig. 7b), deviations between the analytical predictions and numerical simulations become more pronounced, particularly at early times. This discrepancy arises from the breakdown of the linear assumptions in the analytical model. For highly elongated droplets, the interface exhibits large curvature variations and significant geometric nonlinearity, leading to strong coupling between different deformation modes. In addition, the capillary pressure is no longer well approximated by a linear perturbation of the spherical curvature, and the relaxation dynamics involve large-scale fluid redistribution rather than independent modal decay. Despite these limitations, the analytical solution still captures the overall trend of the relaxation process remarkably well, especially at later times when the droplet shape approaches a sphere and the deformation amplitude becomes small. This indicates that the late-stage dynamics are effectively governed by the same linear mechanisms described by the analytical theory.

Interestingly, for large initial elongation, the agreement between simulation and theory is generally better for viscoelastic droplets than for Newtonian droplets, particularly after the initial transient phase. This can be understood from the role of stress relaxation in smoothing the dynamics. In viscoelastic fluids, the finite relaxation time $t_c$ introduces a temporal memory that effectively damps rapid variations in the flow and reduces the influence of high-frequency deformation modes. As a result, the system evolves more gradually toward configurations where the assumptions of the linear theory become valid. In contrast, Newtonian droplets respond instantaneously to curvature gradients, leading to stronger nonlinear effects and more pronounced deviations from the linear analytical predictions during the early stages of relaxation. Thus, viscoelasticity not only introduces additional timescales but can also extend the regime of validity of linearized theoretical descriptions by moderating the nonlinear dynamics.

### Detailed comparison of time history of deformation

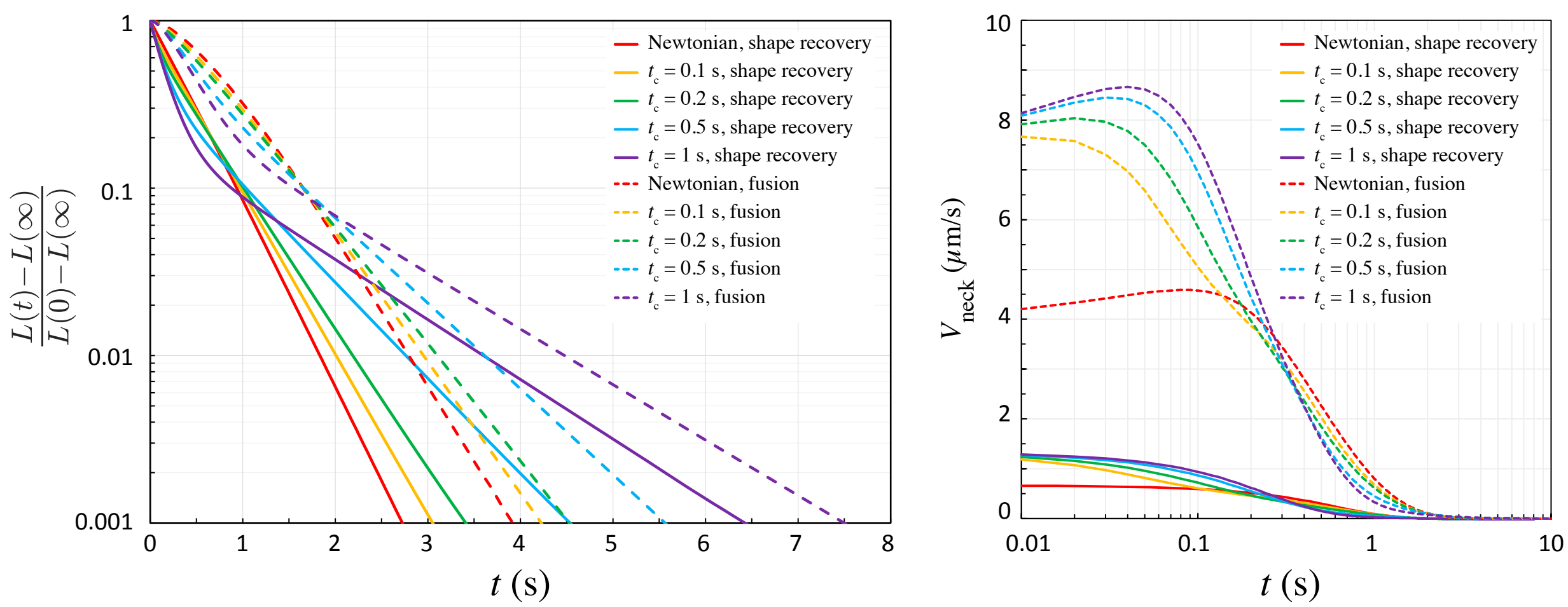

Figure 8: Comparison between shape recovery and fusion dynamics. (a) Time histories of normalized vertical dimension.(b) Velocity of the horizontal dimension. The polymeric viscosity of the droplet is fixed at $\eta_p = 0.25$ Pa·s for both cases and total viscosity is $\eta = 0.5$ Pa·s.

Our FEM simulations demonstrate that, although droplet fusion shares certain qualitative similarities with shape recovery—namely, the ultimate relaxation toward a spherical equilibrium driven by interfacial tension—it is fundamentally a multistage process with richer spatiotemporal dynamics. This increased complexity arises from the initial presence of two disconnected interfaces, the formation and growth of a liquid bridge, and the strong localization of curvature and stress near the neck region during coalescence. As a result, the fusion process cannot be described by a single global relaxation mode, but instead involves a sequence of distinct dynamical regimes governed by different balances of capillary, viscous, and viscoelastic stresses.

In Fig. 8a, we plot the time evolution of the vertical length scale for both shape recovery and fusion. For Newtonian droplets (red solid and dashed lines), the shape recovery process exhibits a straight line on a semi-logarithmic plot, consistent with a single-exponential decay governed by the viscocapillary timescale $\tau_{vc} \sim \eta R/\gamma$. This reflects the fact that shape recovery is dominated by global relaxation of low-order deformation modes. In contrast, the fusion process exhibits clear deviations from single-exponential behavior, with multiple distinct regimes evident in the temporal evolution. At early times, the slope is relatively small, indicating slow initial evolution despite the presence of large curvature at the point of contact. This is followed by a transition to a faster relaxation regime, and ultimately a late-time regime whose slope closely matches that of shape recovery. The convergence of slopes at late times indicates that, once the droplets have fully merged and the interface becomes smooth, the dynamics reduce to a global viscocapillary relaxation similar to that of a single droplet. Notably, this final-stage behavior is largely insensitive to the relaxation time $t_c$, suggesting that it is governed primarily by large-scale geometry and total viscosity.

The multistage nature of fusion is more clearly revealed in the time history of the horizontal velocity associated with bridge expansion (Fig. 8b). In the early stage, theoretical analysis of viscous coalescence predicts that the bridge radius grows linearly with time, corresponding to a constant velocity driven by a balance between capillary pressure at the neck and viscous resistance. This scaling is confirmed by our simulations, which show a plateau in the velocity curve corresponding to this constant-rate expansion regime. Physically, this behavior arises because the dominant curvature—and hence capillary pressure—is localized at the neck, while the surrounding fluid provides viscous dissipation that limits the rate of growth.

Following this initial regime, the system enters a transitional stage in which the bridge expansion accelerates. This acceleration can be attributed to the increase in the characteristic length scale of the bridge, which reduces curvature gradients and allows a larger volume of fluid to participate in the flow. At the same time, the flow field becomes less localized and begins to involve more global redistribution of fluid. This is then followed by a deceleration phase, as the curvature differences continue to decrease and the system approaches a nearly spherical configuration. In this final stage, the dynamics are again governed by global viscocapillary relaxation, similar to shape recovery, leading to a gradual reduction in velocity.

In contrast to the non-monotonic behavior observed in fusion, the velocity associated with shape recovery remains monotonic in time for both Newtonian and viscoelastic cases. This reflects the absence of topological changes and localized singularities in shape recovery, allowing the dynamics to be described by a smooth decay of deformation modes. For viscoelastic droplets, the fusion dynamics are further modified by the presence of an additional intrinsic timescale, the stress relaxation time $t_c$, which introduces memory effects into the system. When $t_c$ is small (e.g., $t_c = 0.1$ s), the fluid behaves effectively as a Newtonian liquid on the timescale of fusion, and the velocity evolution becomes more monotonic, closely resembling the viscous case. However, as $t_c$ increases, elastic stresses persist over longer times and interact with the evolving flow field, leading to more pronounced acceleration during the intermediate stage. This can be interpreted as the temporary storage and subsequent release of elastic energy, which enhances the driving force for bridge expansion beyond purely viscous dissipation. Consequently, viscoelasticity not only modifies the magnitude of the velocity but also amplifies the multistage character of the fusion process, particularly in regimes where the Deborah number $De = t_c/\tau_{vc}$ is of order unity or larger.

### Effects of exterior domain with finite viscosity

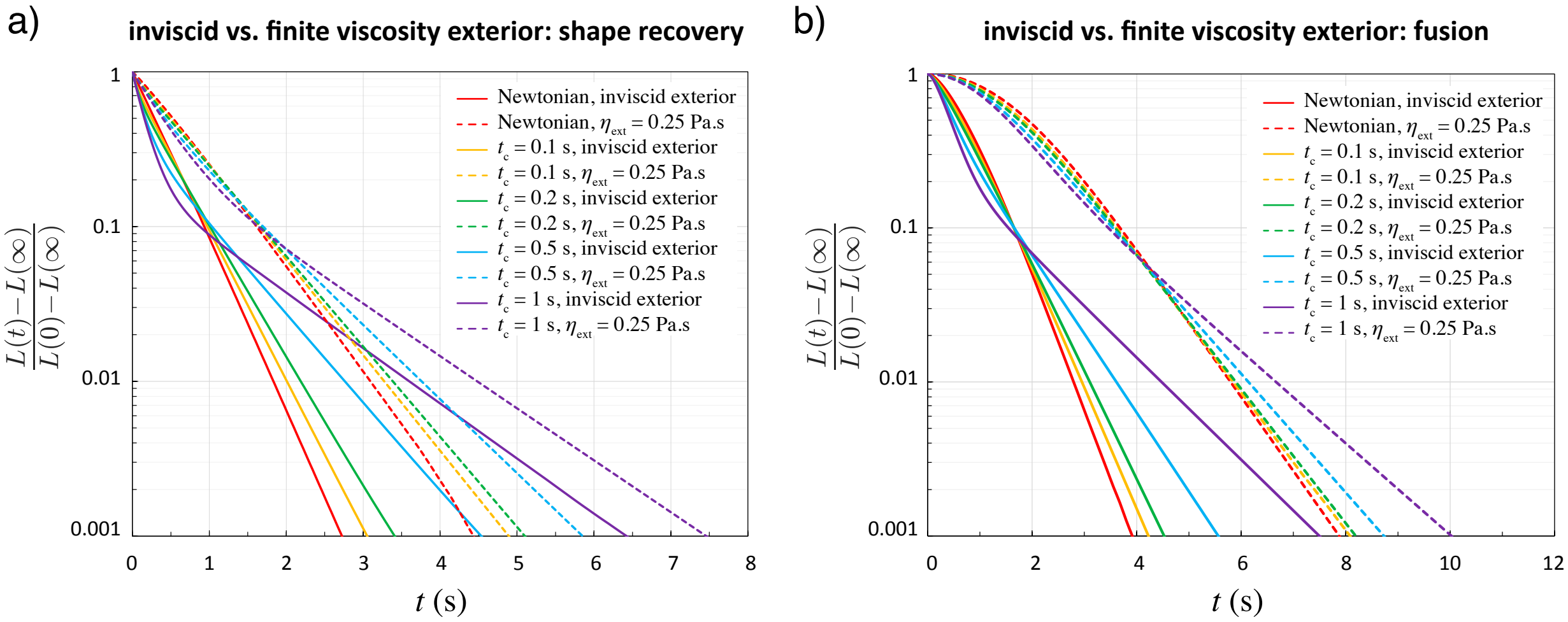

Figure 9: Comparison of the shape recovery and fusion dynamics of two droplets in a inviscid exterior domain and in an exterior domain with finite viscosity, $\eta_{ext} = 0.25\,$Pa·s. The viscosity of the droplet(s) are fixed at $\eta = 0.5\,$Pa·s with the polymeric viscosity fixed at $\eta_p = 0.25\,$Pa·s for all viscoelastic cases.

We then investigated the impact of an exterior fluid domain with finite viscosity on the fusion dynamics and the associated rate of coalescence. To isolate this effect, we compared simulations of droplet fusion in two limiting cases: (i) an effectively inviscid exterior domain, in which viscous dissipation is confined to the interior of the droplets, and (ii) an exterior domain modeled as a Newtonian fluid with finite viscosity $\eta_{ext} = 0.25\,$Pa·s, comparable to the droplet viscosity. In both cases, the droplets themselves are viscoelastic, characterized by a fixed polymeric viscosity $\eta_p = 0.25\,$Pa·s and a range of relaxation times $t_c = 0.1$–$1$ s. The comparison is shown in Fig. 9.

From a physical perspective, droplet fusion is driven by capillary pressure arising from the high curvature at the neck region, while being resisted by viscous and viscoelastic stresses that dissipate or store energy. In the absence of an exterior viscosity, the dominant dissipation occurs within the droplet interior, and the surrounding medium offers negligible resistance to flow. As a result, the flow field induced by capillary forces can extend freely into the exterior, leading to more efficient mass transport and faster bridge expansion.

In contrast, when the exterior domain has finite viscosity, the hydrodynamic problem becomes one of two-phase flow with coupled stress balance across the interface. The interfacial boundary condition now enforces continuity of velocity and a jump in stress proportional to surface tension. In this case, the effective resistance to flow is increased, since the motion of the interface requires deformation not only of the droplet interior but also of the surrounding fluid. This introduces an additional dissipation channel and modifies the overall force balance.

A useful way to interpret this effect is through an effective viscocapillary timescale. In the simplest scaling picture, the characteristic relaxation time for fusion can be expressed as $\tau_f \sim \frac{\eta_{eff} R}{\gamma}$, where $\eta_{eff}$ represents an effective viscosity that accounts for contributions from both the interior and exterior fluids. When $\eta_{eff}$ is non-negligible, $\eta_{eff}$ increases, leading to a longer timescale and thus slower fusion dynamics. This is consistent with classical results for droplet deformation and coalescence in two-fluid systems, where the viscosity ratio between interior and exterior phases plays a central role in determining the rate of interface motion.

The simulation results in Fig. 9 clearly reflect this physical picture. For all values of the relaxation time $t_c$, the presence of a finite-viscosity exterior domain leads to a systematic reduction in the fusion rate, as evidenced by the slower decay of the characteristic length scale. This effect is particularly pronounced during the intermediate and late stages of fusion, where the flow becomes more global and involves a larger volume of the surrounding fluid. In the early stage, where the dynamics are dominated by highly localized curvature at the neck, the influence of the exterior viscosity is comparatively weaker but still noticeable.

For viscoelastic droplets, the interplay between exterior viscosity and stress relaxation introduces additional complexity. While the intrinsic relaxation time $t_c$ governs how quickly elastic stresses within the droplet relax, the exterior viscosity controls how efficiently momentum can be transmitted across the interface into the surrounding fluid. As a result, the overall dynamics are determined by a competition between capillary driving, viscoelastic stress relaxation, and two-phase viscous dissipation. In regimes where $t_c$ is small, the droplet behaves effectively as a Newtonian fluid, and the dominant effect of the exterior domain is to increase dissipation. As $t_c$ increases, elastic stresses persist for longer times and interact with the slowed hydrodynamic response imposed by the exterior fluid, further modifying the temporal evolution of fusion.

Overall, these results demonstrate that the surrounding fluid cannot be neglected when interpreting droplet fusion experiments. Even when the droplet material properties are fixed, variations in the viscosity of the exterior medium can significantly alter the observed fusion dynamics, highlighting the importance of considering two-phase hydrodynamic coupling in both modeling and experimental analysis.

## Empirical formulas for fusion dynamics

Although no closed-form analytical solution currently exists for droplet fusion—either in the purely viscous or viscoelastic regime—due to the intrinsically nonlinear geometry and topological transition involved in coalescence, we propose an empirical representation of the fusion dynamics based on the evolution of a characteristic length scale. Specifically, we fit the normalized reduction in the end-to-end distance using a stretched-exponential form (19):

$$\frac{L(0) - L(\infty)}{2R_0} = 1 - e^{-(t/\tau_f)^\beta} \tag{17}$$

where $\tau_f = 1.97\eta R_0/\gamma$ and $\beta$ is a dimensionless exponent that captures deviations from single-exponential relaxation. Fig. 10 shows that Eq. 17 matches well with the numerical solution. With increasing $t_c$, the curves deviate further from Eq. 17.

From a physical standpoint, this empirical form can be interpreted as a coarse-grained description of a multistage, non-self-similar relaxation process. In contrast to shape recovery—where linear theory predicts a single or a small number of exponential decay modes—droplet fusion involves a continuous evolution of the interface geometry, with different stages governed by distinct dominant balances. In the early stage, the dynamics are controlled by highly localized curvature at the neck, leading to rapid but spatially confined flow. In intermediate stages, the characteristic length scale of the bridge increases, and the flow transitions toward a more global redistribution of mass. Finally, in the late stage, the system approaches a nearly spherical shape, and the dynamics resemble classical viscocapillary relaxation.

The stretched-exponential form effectively captures this continuous crossover between regimes by introducing a distributed spectrum of relaxation times, rather than a single characteristic timescale. When $\beta = 1$, the expression reduces to a simple exponential decay, consistent with single-mode relaxation. However, for $\beta \neq 1$, the relaxation reflects a superposition of processes occurring over multiple timescales. In particular, values of $\beta > 1$ correspond to accelerated dynamics at intermediate times, which is consistent with the observed transition from slow initial bridge formation to faster global relaxation. Thus, the exponent $\beta$ provides a compact way to encode the nontrivial temporal structure of the fusion process.

The characteristic timescale $\tau_f \sim (\eta/\gamma)R_0$ arises naturally from a balance between capillary pressure, which scales as $\gamma/R_0$, and viscous stress, which scales as $\eta U/R_0$, yielding a velocity scale $U \sim \gamma/\eta$ and hence a timescale $\tau \sim R_0/U \sim (\eta/\gamma)R_0$. The prefactor (1.97) reflects geometric and dynamical details of the coalescence process that are not captured by simple scaling arguments but are obtained from fitting to numerical data.

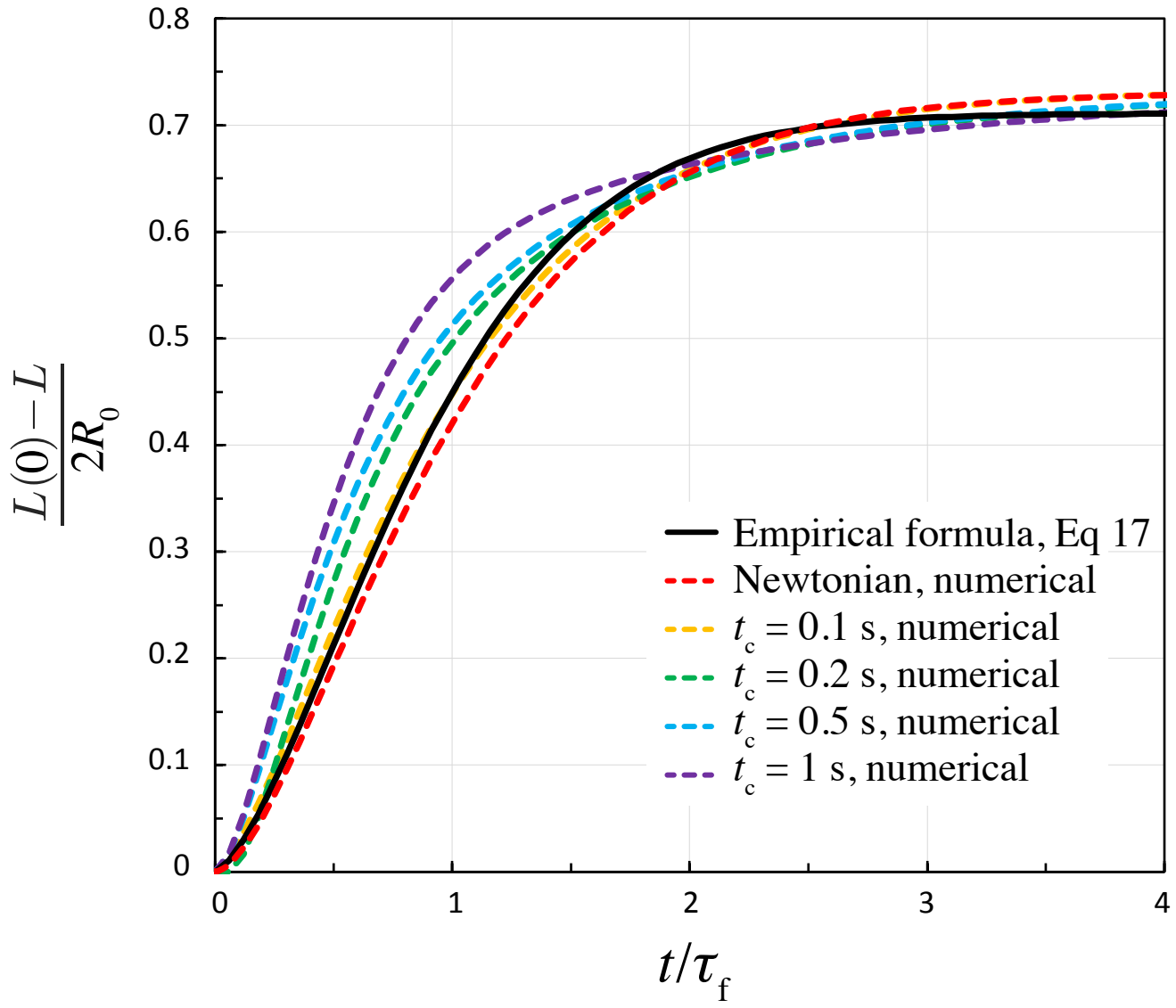

Figure 10: Comparison of the computationally predicted relaxation of droplet fusion with the empirical formula. The dependence of the shrinkage in edge-to-edge distance, $(L(0) - L)/2R_0$, on the dimensionless time $t/\tau_f$ where $\tau_f = 1.97 \left(\frac{\eta}{\gamma}\right) R_0$ . Dashed lines represent the numerical solution; the solid curve is the stretched-exponential fit with $\beta = 1.5$.

As shown in Fig. 10, this empirical expression provides an excellent fit to the numerical results for Newtonian droplets, indicating that despite the absence of an exact analytical solution, the dominant physics can be effectively captured by a single timescale and a stretched-exponential functional form. However, as the viscoelastic relaxation time $t_c$ increases, systematic deviations from Eq. 17 become apparent. This deviation can be attributed to the emergence of an additional intrinsic timescale associated with stress relaxation, which modifies the balance between capillary driving and resistance. In viscoelastic systems, the effective response depends not only on the instantaneous viscosity but also on the history of deformation, leading to temporal memory effects that cannot be fully captured by a single-parameter scaling.

In particular, when the Deborah number $De = t_c/\tau_f$ becomes non-negligible, elastic stresses persist during the fusion process and alter both the rate and temporal profile of relaxation. This results in deviations from the stretched-exponential form derived for purely viscous systems, especially in the intermediate regime where both capillary forces and elastic stresses are significant. These observations suggest that a generalized empirical model for viscoelastic fusion must incorporate additional parameters or modified functional forms to account for the interplay between viscocapillary dynamics and stress relaxation.

## CONCLUSION

In this work, we developed a unified theoretical and computational framework to investigate the dynamics of droplet shape recovery and fusion in Newtonian and viscoelastic fluids. By combining analytical solutions for small-deformation recovery with finite-element simulations of fully nonlinear dynamics, we systematically compared these two processes under comparable physical conditions and identified both their shared mechanisms and fundamental differences.

Our results show that shape recovery is governed by global viscocapillary relaxation of a single connected interface and can be well described by modal decay theory in the small-deformation regime. In contrast, droplet fusion is intrinsically a multistage and multiscale process, initiated by a localized curvature singularity at the point of contact, followed by neck growth and eventual transition to global relaxation. This geometric and dynamical complexity leads to non-exponential behavior and prevents a direct extension of analytical recovery theory to fusion.

We demonstrated that viscoelasticity plays a central role in both processes by introducing an additional intrinsic timescale associated with stress relaxation. The competition between the viscocapillary timescale and the relaxation time governs the transition from Newtonian-like behavior to strongly viscoelastic dynamics. In shape recovery, this results in multi-exponential relaxation and improved agreement with linear theory at later stages. In fusion, viscoelasticity amplifies intermediate-stage dynamics through the storage and release of elastic energy, leading to pronounced deviations from purely viscous scaling, particularly at finite Deborah numbers.

Furthermore, we showed that the surrounding fluid can be important: the presence of an exterior domain with finite viscosity introduces additional hydrodynamic resistance through two-phase coupling, effectively increasing the viscocapillary timescale and slowing fusion dynamics. This highlights the importance of considering both internal rheology and external hydrodynamic conditions when interpreting experimental observations.

Finally, we compared a stretched-exponential empirical formula for fusion dynamics with our numerical solutions. While this formula captures Newtonian fusion behavior well, its breakdown in viscoelastic regimes underscores the need for more general theoretical descriptions that incorporate stress relaxation and memory effects.

Overall, this work establishes a mechanistic understanding of droplet remodeling that bridges analytical theory, numerical simulation, and experimental interpretation. It provides a quantitative framework for distinguishing between shape recovery and fusion dynamics and offers practical guidance for extracting viscoelastic material properties of biomolecular condensates from dynamic measurements.

## AUTHOR CONTRIBUTIONS

Z.P. and H.Z. designed the research. M.M.N. performed numerical simulations. M.M.N., Z.P. and H.Z. analyzed the data. All authors contributed equally to writing the article.

## ACKNOWLEDGMENTS

Z.P. and M.M.N. acknowledge the supports from US NSF grants NSF CMMI2339054 and NSF PHY2210366, and Scholar Award from American Society of Hematology. H.Z. acknowledges the support from National Institute of General Medical Sciences under Grant No. GM118091.